\pdfoutput=1

 \documentclass[10pt, conference]{IEEEtran} 

\usepackage[left= 0.625in,right= 0.625in,top=0.7in,bottom=1in]{geometry}

\usepackage{bbm}
\usepackage{amsthm}
\usepackage{amsmath}
\usepackage{amsfonts}
\usepackage{amsmath,amsfonts,amssymb,amsbsy, amsthm,ae,aecompl}
\usepackage{algorithmic, algorithm, float}
\usepackage[english]{babel}
\usepackage{bm}
\usepackage{color}
\usepackage[noadjust]{cite}
\usepackage{epsfig}
\usepackage{enumerate}
\usepackage{float} 
\usepackage{fancyhdr}
\usepackage[T1]{fontenc}
\usepackage[acronym,toc,shortcuts]{glossaries}
\usepackage{graphicx, caption, subcaption}
\usepackage{hyperref}
\usepackage{lastpage}
\usepackage{listings}
\usepackage{lipsum}
\usepackage{dsfont}
\usepackage{multirow,tabularx}
\usepackage[normalem]{ulem}
\usepackage{tikz,pgfplots}
\usepackage{times}
\usepackage{verbatim}
\usepackage[all]{xy}
\usepackage{mathtools}
\usepackage{tikz}
\usepackage{tikz-qtree,tikz-qtree-compat}
\usepackage{pgfplots}
\usepackage{lipsum}
\usetikzlibrary{calc}
\usetikzlibrary{patterns}

\usepackage{enumitem}
\graphicspath{{figures/}}

\usetikzlibrary{arrows,shapes}
\newcommand{\E}[1]{{\mathbb E}\left[ #1 \right]}

\hypersetup{
	colorlinks,%
	citecolor=black,%
	filecolor=black,%
	linkcolor=black,%
	urlcolor=black
}

\pgfplotsset{
	grid style = {
		dash pattern = on 0.025mm off 0.95mm on 0.025mm off 0mm, 
		line cap = round,
		black,
		line width = 0.5pt
	},
	tick label style={font=\small},
	label style={font=\small},
	legend style={font=\footnotesize},
}

\pgfplotscreateplotcyclelist{laneas1}{
	cyan!60!black,		solid, 	every mark/.append style={fill=cyan!60!black},mark=x\\%
	cyan!60!black,		solid, 	every mark/.append style={fill=cyan!60!black},mark=+\\%
	cyan!60!black,		solid, 	every mark/.append style={fill=cyan!60!black},mark=o\\%
	red!80!black,       dashed\\%
	cyan!60!black, 		densely dotted,	every mark/.append style={fill=cyan!80!black},mark=diamond*\\%
}

\newcommand{\bfW}{ \mathbf{W} }
\newcommand{\bfx}{ \mathbf{x} }
\newcommand{\bfq}{ \mathbf{q} }
\newcommand{\bfy}{ \mathbf{y} }

\newcommand{\trace}{ {\mathrm{Tr}} }
\newcommand{\diag}{ \mathrm{diag} }

\definecolor{bluegreen}{rgb}{0.0, 0.87, 0.87}
\definecolor{bluencs}{rgb}{0.0, 0.53, 0.74}

\ifCLASSINFOpdf
\else
\fi

\newtheorem{lemma}{Lemma}

\newtheorem{Definition}{Definition}

\IEEEoverridecommandlockouts

\begin{document}
	\title{Enhancing Favorable Propagation in Cell-Free Massive MIMO Through Spatial User Grouping }
	
	\author{\IEEEauthorblockN{$\text{Salah Eddine Hajri}^*, ̃ \text{ Juwendo Denis}^*, 
			̃ \text{and Mohamad Assaad}^*$}\\
		\IEEEauthorblockA{*TCL Chair on 5G, Laboratoire des Signaux et Systemes (L2S, CNRS), CentraleSupelec	Gif-sur-Yvette, France\\
			\{Salaheddine.hajri,\; juwendo.denis,\; Mohamad.Assaad\}@centralesupelec.fr}
			\thanks{ This research has been performed in the framework of the Horizon 2020 project ONE5G (ICT-760809) receiving funds from the European Union. }
		
	}
	
	
	%

	
	\maketitle
	
	\begin{abstract}
       Cell-Free (CF) Massive multiple-input multiple-output (MIMO) is a distributed antenna system, wherein a large number of back-haul linked access points randomly distributed over a coverage area  serve simultaneously  a smaller number of users. CF Massive MIMO inherits  favorable propagation  of  Massive MIMO systems. However,  the level of favorable propagation which highly depends on the network topology and environment may be hindered by user' spatial correlation.  In this  paper, we investigate the impact of the network configuration on  the level of  favorable propagation for a  CF Massive MIMO network.  We formulate a user  grouping  and  scheduling  optimization problem  that  leverages  users' spatial diversity. The formulated design optimization problem  is proved  to be NP-hard in general. To circumvent the prohibitively high computational cost, we adopt the semidefinite relaxation  method to find a sub-optimal solution. The effectiveness of the proposed strategies  is then verified through numerical results which demonstrate a non-negligible improvement in the performance of the studied scenario. 
	\end{abstract}
	
	\IEEEpeerreviewmaketitle
	\vspace{-0.2cm}
	\section{Introduction}    

Recently, cell-free (CF) massive multiple-input multiple-output (MIMO) systems have attracted a lot attention \cite{CF_LSFD,CF_SC, Nayebi_TWC} and   have been recognized as an effective and appealing approach for next generation wireless networks. CF massive MIMO systems consist of very large number of single-antenna distributed access-points (APs) serving simultaneously,  over same time/frequency resources,  a relatively small number of users \cite{CF_SC}. CF massive MIMO  improves considerably macro-diversity and  provides the network with substantially higher coverage probability.  
	  
	 CF massive MIMO can rely on channel reciprocity and  uplink  training in order to acquire  channel  state information  (CSI) by adopting time-division duplex (TDD) mode  \cite{CF_SC}. 
With a large number of  APs, CF massive MIMO  can exploit  favorable propagation that results from mutual orthogonality of users' channel \cite{fp_massive}.  In \cite{CF_SC}, the authors leveraged favorable propagation to derive  closed-form expressions for the downlink  and uplink achievable rates  for CF massive MIMO systems.  The  spatial correlation resulting from  the distributed deployment of APs may however have a detrimental impact  on the favorable propagation. More specifically, users that are relatively closed to each other will incur high spatial correlation which will jeopardize the mutual orthogonality of the users' channel. 

 Channel  hardening  and favorable  propagation  in CF massive MIMO  using stochastic geometry model  were investigated in \cite{DBLPzheng}. Therein, the authors concluded that  one may not completely rely on channel hardening and favorable propagation when
assessing  the system' performance since the derived bounds may not be tight  due to the  impact of spatial correlation between some users.  Consequently, improving favorable propagation in presence of distributed antennas systems is of great necessity. 
		 
 In this  work,   we  analyze how  spatial correlation between users' channels vector influences  favorable propagation and explore how to improve orthogonality between users' channel by taking into account solely the large-scale fading and the number of  available APs.
We establish a design optimization problem based on the perspective of user's scheduling. The proposed design advocates  how users' grouping can improve favorable propagation. However, the resulting optimization problem is difficult to solve in general. We demonstrate that the  formulated problem is NP-hard and  we invoke semidefinite relaxation (SDR) \cite{Luo_SDR}  approach to design a polynomial time solvable randomized procedure to find a sub-optimal solution to the NP-hard problem.  In addition to that, to increase users' throughput, we investigate the problem of  bandwidth allocation for the resulting scheduling design strategy.  The numerical results demonstrate that considerable performance improvement is achieved through the proposed efficient user grouping procedure and the bandwidth allocation scheme.
		   
\vspace{-0.03cm}
	
	\section{System Model And Preliminaries}
	
	We consider a CF massive MIMO  system  that consists of of $K$ single omni-directional antenna users that are served simultaneously by $M$ single antenna APs. In this work, It is assumed that $K<<M$ and that the APs are using the  same  time-frequency  resources. APs are randomly located within a given  coverage area and  are managed by a central processing unit (CPU)  to  which  they are connected  through  perfect back-haul links.
	The CPU  handles  part of  the physical layer processing such as data coding and decoding. 
	 Let $\bm{g}_{k} \in \mathbb{C}^{M \times 1}$,  denote the complex channel vector between user $k$ and all the APs. Specifically, the $m$-th  element,   $g_{mk}$ is the channel coefficient between the $k$-th user and the  $m$-th AP and is modeled as follows: \vspace{-0.1cm}
		\begin{align}\label{eq:channel_element}
		& g_{mk} = \sqrt{\beta_{mk}} h_{mk }
		\end{align}
	where $ h_{mk } \sim \mathcal{C}\mathcal{N}(0,1), m=1,\cdots,M,\quad k=1,\cdots,K,$ denote the small-scale fading coefficients which  are independent and identically distributed (i.i.d) while  $ \beta_{mk }, m=1,\cdots,M,\quad k=1,\cdots,K,$  the large scale fading coefficients that include path-loss and shadowing. 

	Similarly to the work done in \cite{CF_SC}, we assume uplink/downlink  channels reciprocity. 
Provided that the system operates according to a TDD protocol,  each coherence interval is divided between  uplink training,  downlink and uplink data  transmission. In  order to   perform  multiplexing and  de-multiplexing, the APs need to acquire CSI  through  uplink training.  
Let $T_c$ be the length of the  coherence interval (measured in samples) and  $\tau$,  the uplink training duration with $\tau < T_c$.  During the  uplink  training  phase, a maximum of  $\tau$ users simultaneously send  their pilot  sequences   to the APs. Under such setting,     we consider a set of orthonormal training sequences, denoted by $\bfq_k\in \mathbb C^{\tau \times 1}, \forall k=1, \cdots, \tau$ such that $ \bfq^{\dagger}_k \bfq_j =\delta_{kj} $ (with $\delta_{kj}$ be the Kronecker delta).  
	
Channel estimation is  performed in  a decentralized fashion where each AP $m$ independently estimates $g_{mk}$ of  the  $\tau$ active users. The received pilot vector signal at  the $m$th AP can be expressed as \vspace{-0.3cm}
 		\begin{align}\label{eq:uplink_training}
 		& \bm{y}_{m,p}=   \sqrt{\rho_p} \sum\limits_{k=1}^{\tau} g_{mk} \bfq_k + \bm{n}_{m,p}
 		\end{align} 
  where $\rho_p$ is the transmit power during the training phase  and $\bm{n}_{m,p}$ is the  additive Gaussian noise vector at the $m$-th AP. The elements of  $\bm{n}_{m,p}$  are i.i.d. random variables.
  The $m$-th AP  performs minimum mean-square error  (MMSE) channel estimation  using $\bm{y}_{m,p}$  in order to  obtain the channel estimates  given by \vspace{-0.3cm}
	\begin{align}\label{eq:channel_estimates}
	& \hat{g}_{mk}=   \frac{\sqrt{\rho_p}  \beta_{mk}   }{  \rho_p  \beta_{mk}       +1} \big( \sqrt{\rho_p}  g_{mk}  + \bm{n}_{m,p} \bfq^\dagger_k \big),  \,   k=1,\cdots,\tau 
	\end{align}
Since  the pilot sequences are mutually  orthogonal, there  is no  pilot contamination and  the channel  estimate of each user  $k$ is independent of    $g_{mj}, \forall j \neq k$. 
Each AP  independently  estimates the channel for each active user  and  the channel estimate is then used to precode the downlink signal. As in \cite{CF_SC}, we assume that conjugate beamforming is employed to communicate with the active users. Consequently, the transmit signal  of the  $m$-th AP can be expressed as \vspace{-0.3cm}
	\begin{align}\label{eq:downlink_ap}
	& y_{m,d}= \sqrt{\rho_d} \sum\limits_{k=1}^{\tau} \hat{g}^*_{mk} d_k
	\end{align}
where $d_k$ with  $\E{\lvert d_k\rvert^2}=1$,  denotes  the data symbol intended to user $k$ while $\rho_d$, the  downlink transmit power. The  received signal at the $k$-th user  is then given by 
\begin{align}\label{eq:dsignal_k_user_decomposition}
	& r_{k,d}= \sqrt{\rho_d} \left(  \sum\limits_{m=1}^{M} {g}_{mk} \hat{g}^*_{mk} d_k+	
	 \sum\limits_{m=1}^{M} \sum\limits_{j \neq k}^{\tau}  {g}_{mk} \hat{g}^*_{mj} d_j \right)+ n_{k,d},	 
	\end{align}
where $n_{k,d}$ denotes the white additive Gaussian noise at user $k$. 
By assuming  a large number of APs, users can detect their downlink data using only the channel statistics \cite{CF_SC}. The resulting  rate  of a given user $k$  can be written as \cite{EE_CF}
	{\small{\begin{align}\label{eq:dsignal_k_user_decomposition}
	& R_{k,d}= \mathcal{B} \big(1-  \frac{\tau}{T_c} \big)\; \text{log}_2 \left(	1+\frac{\rho_d (\sum\limits_{m=1}^{M} \nu_{mk}     )^2   }{\rho_d\sum\limits_{m=1}^{M} \sum\limits_{j =1}^{\tau}  \nu_{mk'} \beta_{mk} +1}  \right)
	\end{align}}}where $\mathcal{B}$ denotes the  bandwidth of the system and $ \nu_{mk}\triangleq  \frac{\rho_p\beta_{mk}^2}{1+\rho_p\beta_{mk}} $  is the  variance of $\hat{g}_{mk}$, the MMSE estimate of the channel.

	\section{ Improving favorable propagation:  which users can be active simultaneously?}
	
	Favorable propagation  represents an important property in   large antenna systems. It refers to  the  mutual  orthogonality between users' vector wireless channel.
	With favorable propagation, the overall system' performance is guaranteed to be very appealing with simple linear processing \cite{fp_massive} since the effect of interferences is considerably attenuated.  
	To  obtain favorable propagation, users' channel vectors need to  be mutually orthogonal, i.e 
				\begin{equation}\label{eq:favorable}
				\begin{aligned}
				&  \bm{g}^\dagger_k \bm{g}_j=
				\left\{
				\begin{array}{ll}
				0 & \mbox{if $k\neq j$,}  \\
			     \lVert \bm{g}_k\rVert^2  \neq 0& \mbox{otherwise, } 
				\end{array}
				\right.\\
				\end{aligned}
				\end{equation}	
	Practically, the condition in \eqref{eq:favorable} cannot be exactly met, but it can be approximately  achieved. This is the case when
	 the number of antennas  grows large and the  channels are said to provide  asymptotically favorable propagation. The asymptotically favorable propagation condition is  stated as
	 	\begin{align}\label{eq:favorable_asympt}
	 	&     \frac{\bm{g}^\dagger_k \bm{g}_j}{M}      \longrightarrow  0,\; M  \longrightarrow\infty  \quad  \text{for} \quad k \neq j,  
	 	\end{align} 
	By using the channels gain from each AP to the users, equation \eqref{eq:favorable_asympt}  can be  equivalently rewritten  as 
		\begin{align}\label{eq:favorable_asympt_product}
		&     \frac{ \sum\limits_{m=1}^{M} \sqrt{\beta_{m,k}}  \sqrt{\beta_{m,j}} h^*_{m,k} h_{m,j}}{M}      \longrightarrow  0  \quad  \text{for} \quad k \neq j,  
		\end{align}
	Contrarily to collocated massive MIMO systems,  the large scale fading coefficients from  each AP to  a given user are different in a cell-free Massive MIMO network.  Intuitively, this  spatial diversity will  have a considerable impact on  the favorable propagation.
	 Consequently, spatial channel correlation is an important parameter that need to be taken into consideration  to  improve favorable propagation in the system.

	In a practical scenario,   the  number of APs can be made very  large but cannot grow indefinitely. Provided that asymptotically favorable propagation is achieved when $M$ tends to infinity, the condition in \eqref{eq:favorable_asympt} will not meet for practical scenario.  Given that favorable propagation is important to achieve good performance of CF massive MIMO, we use a different perspective to make condition in \eqref{eq:favorable_asympt} works for practical scenario.  To do so, we leverage  the complementary  cumulative distribution function  of the  inner product  between two given users'  channel \vspace{-0.5cm}
	   \begin{align}\label{eq:favorable_asympt_prob}
	   &   P_\theta= \mathbb{P} \left\{\frac{\bm{g}^\dagger_k \bm{g}_j}{M}  \geq \theta    \right\}  
	   \end{align} 
 Concretely,  to  improve favorable propagation, $P_\theta$ should be as close to zero as possible for any value of  $\theta \geq 0$.  Making  $P_\theta, \;\forall \theta \geq 0$ arbitrarily small  means that the users' channel vectors achieve \emph{near orthogonality}. 
   Using  Chebychev's inequality \cite{prob_ineq},  $P_\theta$ can be lower-bounded  by 
   \begin{align}\label{eq:favorable_inequality}
   &   P_\theta= \mathbb{P}\left \{  \frac{\bm{g}^\dagger_k \bm{g}_j}{M}  \geq \theta   \right\}    \leq  \frac{1}{1+ \frac{M^2 \theta^2}{\sum\limits_{m=1}^{M} {\beta_{m,k}}  {\beta_{m,j}}}} ,  
   \end{align}
	 From~\eqref{eq:favorable_inequality}, it can be observed that a viable way to reduce the value of $P_\theta$,  is  to minimize the inner product  between the users large scale  fading  vectors i.e., $ \sum\limits_{m=1}^{M} {\beta_{m,k}}  {\beta_{m,j}}$ which quantifies the  spatial  correlation  between the channels  of users  $k$ and $j$. 	Consequently, reducing spatial  correlation  between active  users' channel  in a CF massive MIMO  system can considerably improve favorable propagation.

 One plausible way to reduce spatial correlation is to resort to appropriate selection of active users. In this  work, we construct a user' scheduling optimization problem that  enables  users to be active simultaneously  only when their  channels have low  spatial correlation.   Our proposed scheme is discussed in the next section.
\vspace{-0.5cm}
	 \subsection{Graphical Modeling and proposed solution}
	In the  considered setting,  the first step is  to construct a spatial  correlation graph that  captures the level of  favorable  propagation  for a set of users which are active  simultaneously.  
	More specifically,   we design an undirected favorable propagation graph $\mathcal{G}(\mathcal{V},\mathcal{E})$. The set of of vertices  $ \mathcal{V} $ represents the users  in the  coverage  area. Each edge  $e_{k,j} \in \mathcal{E}$ is associated with  a weight $\omega_{k,j} \triangleq  \sum\limits_{m=1}^{M} \beta_{mk} \beta_{mj}, $ which is directly related to the spatial  correlation between the two users channel.
    Using  the constructed graph $\mathcal{G}(\mathcal{V},\mathcal{E})$, we  formulate a user selection optimization  problem. 
   
    The considered optimization consists of minimizing the  spatial correlation between the channels  of  users that belong to the same group. Consequently, we formulate a problem where we maximize the inter-group weights. The latter is equivalent to  constructing  groups with improved favorable propagation since users having high spatial correlation between their channels will be allocated to different groups.
   It is worth mentioning that  the resulting  uplink training  overhead should be taken into account in the formulated optimization problem. To this end, the cardinality of each group should not violate a certain threshold  so that CSI  estimation can be performed without  pilot contamination  with a maximum  training sequence of length  $\tau$. Moreover,  to fully exploit spatial  diversity, a  given user  is allowed to belong to multiple  groups simultaneously.
    
Define the following variable
    \begin{equation}
    \begin{aligned}
    & x_{k,c}   =
    \left\{
    \begin{array}{ll}
    1  & \mbox{if  user $k $ is allocated to the $c$-th group}  \\
    0 & \mbox{otherwise} 
    \end{array}
    \right.\\
    \end{aligned}
    \end{equation} The user scheduling problem is formulated  as the following  combinatorial optimization problem
             \begin{equation}\label{eq:problem_grouping}
			\begin{aligned}
		\max_{\substack{ x_{k,c}  \in \{0, 1\} , \,\forall c, \forall k}}~& \sum_{c=1}^{C}  \sum_{k \in\mathcal{V} }^{} \sum_{j \in\mathcal{V}, j \neq k}^{}  w_{k,j}\left( 1- x_{k,c} \right)x_{j,c}\\
		\text{s.t.} 	~&  \sum_{c=1}^{C} x_{k,c} \leq \alpha, \forall k \in  \mathcal{V}    ,\\
			~&  \sum_{k \in\mathcal{V}}^{}  x_{k,c} \leq \tau  , \forall c=1,\ldots,C, 
			\end{aligned}
               \end{equation}
where $C$ denote the total number of groups and $\alpha$, the maximum number of groups to which a user can belong at the same time.  

Before proceeding to solve problem \eqref{eq:problem_grouping}, we  investigate its  computational tractability for $\alpha=1$. This is done through the following lemma.  
			\begin{lemma}  
			Problem \eqref{eq:problem_grouping} is  NP-hard in general.
			\end{lemma}
	\emph{Proof:}  		We demonstrate the NP-hardness of problem \eqref{eq:problem_grouping} by considering a special case of our setting. Specifically, we study the complexity of problem \eqref{eq:problem_grouping} for $\alpha=1$ and  $C*\tau \geq K$. It refers to the case where  each user is allocated,  at most, to one group and the constraints allow each users to be allocated at least once.
The goal is to build the equivalence between this special case and the problem of partition into cliques of bounded size which is defined as follow \cite{hard1}: 		
		\begin{Definition}\label{Definition_problem}
					Consider a graph $\mathcal{G}(\mathcal{V},\mathcal{E})$, a set function $f : 2^\mathcal{V} \longrightarrow R^+$ and a bound $\tau \in  Z^+$. The
					problem is to find a partition of the graph $\mathcal{G}$ into cliques $K_1,..., K_C$ of size at most $\tau$, that is,$ \lvert K_c\rvert \leq \tau, c = 1,..., C$, such that the objective function $\sum_{c=1}^{C} f(K_c)$ is minimized.
				\end{Definition}			
		For $\alpha=1$, the maximization of the objective function of \eqref{eq:problem_grouping} is equivalent to maximizing $ \sum_{c=1}^{C}  \sum_{k \in\mathcal{V} }^{} \sum_{j \in\mathcal{V}, j \neq k}^{}  w_{k,j}  -\sum_{c=1}^{C}  \sum_{k \in\mathcal{V} }^{} \sum_{j \in\mathcal{V}, j \neq k}^{}  w_{k,j} x_{k,c}x_{j,c}$.
		As $C*\tau \geq K$, each user will be allocated to a given cluster. The first sum is then nothing but $2$ times the total weight between users. 
		The considered optimization is then equivalent to minimizing $\sum_{c=1}^{C}  \sum_{k \in\mathcal{V} }^{} \sum_{j \in\mathcal{V}, j \neq k}^{}  w_{k,j} x_{k,c}x_{j,c}$.
			Consequently, the simplified  setting, with the cardinality constraints per group,  is equivalent to solving  a cardinality constrained graph partitioning into cliques  that minimizes the sum of  the clique functions given here by $\sum_{c=1}^{C} f(K_c)= \sum_{c=1}^{C}  \sum_{k \in\mathcal{V} }^{} \sum_{j \in\mathcal{V}, j \neq k}^{}  w_{k,j} x_{k,c}x_{j,c}$.

			In graph theory,  a clique is a subset of vertices of an undirected graph with complete induced subgraph. 
					The aim of the problem given in Definition (\ref{Definition_problem}) is to partition the graph into cliques $K_1..K_{C}$ of maximum size $\tau$ such that
					the sum of the cost function over the cliques is minimized. The cost function in our case is given by the sum of the edges'  weights  that have both their endpoints in the same  cliques.
					Cardinality constrained graph partitioning into cliques with cost minimization contains  the	classic NP-hard clique cover
					problems. It is known as an NP-hard  problem even with a submodular cost function  in complete graphs \cite{hard1}. Considering that problem \eqref{eq:problem_grouping} is equivalent to solving a cardinality constrained graph partitioning into cliques with cost minimization on the complete graph of spatial correlation,   we  deduce that  it  is a NP-hard problem.
							
				One can prove the NP-hardness  of \eqref{eq:problem_grouping} by another method. In fact, we can show that for $\alpha= 1$, \eqref{eq:problem_grouping} is equivalent to another NP-hard problem, namely the capacitated max-$C$-cut problem. We skip the details for brevity.

					\hfill{$\blacksquare$} 

Since problem \eqref{eq:problem_grouping} is NP-hard for alpha=1,  its global optimal solution cannot be found by mean of  polynomial time solvable algorithms. For the general case of problem \eqref{eq:problem_grouping},  where $\alpha \neq 1$, we design here a low-complexity algorithm to find a local optimal solution to problem \eqref{eq:problem_grouping}. To do so, we will resort to semidefinite programming \cite{Vandenberghe_1996}.  	
		
Define following variables and changes of variables
                       \begin{equation}\label{eq:variables}
			\begin{aligned}
    \bfx_c ~& \triangleq \left(x_{1,c} \cdots, x_{K,c}   \right)^\top, \, \bfy_c \triangleq 2\bfx_c-\mathbf{1}_K \\
 \mathbf{W} ~& \triangleq
\begin{pmatrix}
 0 & w_{2,1}& \cdots & w_{K,1}  \\
w_{1,2} & 0& \cdots & w_{K,2}  \\
\vdots & \vdots& \ddots & \vdots  \\
w_{1,K} & w_{2,K} & \cdots & 0  \\
\end{pmatrix}
			\end{aligned}
                       \end{equation}
where $\mathbf{1}_K$ is a column vector which entries are $1$.  Using \eqref{eq:variables}, problem \eqref{eq:problem_grouping} can be equivalently reformulated as 
\begin{subequations}\label{prob:optim_equi}
\begin{align}
\max_{\substack{ }}~&\frac{1}{4} \sum_{c =1}^{C}\left( \varsigma -\bfy_c^\top \bfW  \bfy_c  \right)   \\
\text{s.t.} ~& \sum_{c=1}^C \bfy_c \leq \bar{\alpha} \label{eq: constra_1}\\
~& \trace\left( \diag\left(\bfy_c   \right)  \right) \leq \bar{\tau}, \, \forall c  \label{eq: constra_2}\\
~&\bfy_c  \in \{-1,1 \}^K, \forall c
\end{align}
\end{subequations}
where $\varsigma =\mathbf{1}_K^\top\bfW\mathbf{1}_K, \,   \bar{\alpha}= 2\alpha-C, \, \bar{\tau}= 2\tau-K$. The proposed method consists of combining the semidefinite relaxation method \cite{Luo_SDR} with the \emph{Schur complement} \cite{Stephen2004}. More specifically,  the quadratic terms $\bfy_c^\top \bfW  \bfy_c= \trace\left(\bfW  \bfy_c \bfy_c^\top \right), \, \forall c $ of the optimization problem \eqref{eq:problem_grouping}  is approximated by the linear terms $\trace\left( \bfW  \mathbf{\text{Y}}_c \right), \, \forall c  $ with the rank-one matrices $ \bfy_c\bfy_c^\top, \, \forall c$ being replaced by  positive semidefinite matrices  $\mathbf{\text{Y}}_c, \, \forall c$ of arbitrary rank.   The approximated problem is therefore formulated as 
\begin{equation}\label{prob:optim_sdr_approximation}
\begin{aligned}
\max_{\substack{ \mathbf{\text{Y}}_c \succeq  0, \,\forall c }}~&\frac{1}{4} \sum_{c =1}^{C}\left( \varsigma -\trace\left( \bfW  \mathbf{\text{Y}}_c \right)  \right)   \\
\text{s.t.} ~& \sum_{c=1}^C \bfy_c \leq \bar{\alpha} \\
~& \trace\left( \diag\left(\bfy_c   \right)  \right) \leq \bar{\tau}, \, \forall c \\
~& \diag\left(\mathbf{\text{Y}}_c  \right)=\mathbf{1}_K \\  
\quad~&
\begin{pmatrix}
\mathbf{\text{Y}}_c& \bfy_c  \\
\bfy_c^\top & 1
\end{pmatrix}
\quad \succeq  0, \, \forall c  \\
\end{aligned}
\end{equation}
Problem \eqref{prob:optim_sdr_approximation} is a standard convex optimization problem and can be efficiently solved using interior-point based solvers such as CVX \cite{cvx}. Since the optimization problem \eqref{prob:optim_sdr_approximation} is a relaxation of problem \eqref{prob:optim_equi}, the  optimal $\mathbf{\text{Y}}_c^\star, \,\forall c $ may not be of rank one. Hence, we resort to a randomized procedure, in the vein of Gaussian randomization \cite{Luo_SDR}, to convert the optimal solution of \eqref{prob:optim_sdr_approximation} into a feasible solution to problem \eqref{prob:optim_equi}. The proposed randomized scheme is summarized in Algorithm \ref{alg:random_algol}

\begin{algorithm}[H]
\begin{algorithmic}[1]
\STATE {\bf{input}} an optimal solution $\mathbf{\text{Y}}_c^\star, \,\forall c $ to problem \eqref{prob:optim_sdr_approximation}.
\STATE Generate $\bm{\xi}_c \sim \mathcal{N} (\bm{0},\mathbf{\text{Y}}_c^\star ), \, \forall c$;
\STATE Set $\widetilde{\bm{\xi}_c}= \bm{\xi}_c / \trace\left( \diag\left(\bm{\xi}_c  \right) \right), \, \forall c$;
\STATE Generate $L$ vector samples $\widetilde{\bfy}_c^l, \, l=1,\cdots, L$ feasible for problem \eqref{prob:optim_equi} such that each entry $ \widetilde{y}^l_{k,c}, \, k=1,\cdots, C $  is drawn from the following distribution: \vspace{-0.2cm}
 \begin{equation}
    \begin{aligned}
    & \widetilde{y}_{k,c}^l   =
    \left\{
    \begin{array}{ll}
    1  & \mbox{with probability $(1+\widetilde{\xi}_{k,c})/2$}  \\
    -1 & \mbox{with probability $(1- \widetilde{\xi}_{k,c})/2$} 
    \end{array}
    \right.\\
    \end{aligned}
    \end{equation}
\STATE Compute {\small{$l^\star= \arg max_{l=1,\cdots, L}\frac{1}{4} \sum_{c =1}^{C}\left( \varsigma - \left(\widetilde{\bfy}_c^l\right)^\top \bfW  \widetilde{\bfy}_c^l \right) $}};
\STATE {\bf{output}} the solution $\widehat{\bfy}_c= \widetilde{\bfy}_c^{l^\star}, \, \forall c$.
\end{algorithmic}
\caption{A randomized algorithm  to solve  problem  \eqref{prob:optim_equi} }
\label{alg:random_algol}
\end{algorithm}

\subsection{Bandwidth allocation problem}
Once a solution to problem  \eqref{prob:optim_equi} is found, we proceed to investigate the problem of bandwidth allocation. We denote $\Gamma(c)$, the set of users that belong to group $c$.  Each  group will be active  over the assigned spectrum.  The problem is formulated as \vspace{-0.3cm}
      \begin{equation}\label{eq:problem_Bandwidth}
	\begin{aligned}
\max_{\substack{ 0 \leq \gamma_c  \leq 1, \,\forall c }}~&  \sum_{c, k \in{\Gamma}(c)}^{}\gamma_c  \widetilde{R}_{k,c} \\ 
       \text{s.t.} ~&\overline{R}_k \leq \sum_{c=1}^{C}\gamma_c \widetilde{R}_{k,c}, \forall k \\ 
	~&  \sum_{c=1}^{C} \gamma_c \leq 1
       \end{aligned}
  \end{equation}
where ${\small{\widetilde{R}_{k,c} \triangleq \mathcal{B} \frac{T_c- \lvert {\Gamma}(c) \rvert}{T_c} \; \text{log}_2 \left(	1+\frac{\rho_d (\sum\limits_{m=1}^{M} \nu_{mk}     )^2   }{\rho_d\sum\limits_{m=1}^{M} \sum\limits_{ \substack{k' \in \Gamma(c)\\ k' \neq k} }^{\tau}  \nu_{mk'} \beta_{mk} +1}  \right ) }} $ is the rate of  the $k$th user within group $c$. And, $\overline{R}_k$ is the minimum rate requirements for the $k$th user. Problem \eqref{eq:problem_Bandwidth} is a convex linear optimization problem and the optimal solution can be found using the interior-point method \cite{Stephen2004}.

		\section{Numerical results}
		
		 In this section,  numerical  results  are provided in order to assess the performance of  the  proposed schemes. We consider  a circular region having an  area of $1 \; \text{Km}^2$ where $M$ APs and $K$ users are randomly located according to a uniform distribution. 	The  large-scale fading coefficients include the impact of the  path loss, which  is computed using	
	    a three slope  path loss model \cite{EE_CF}, and  log-normal shadow fading with standard deviation $\sigma_{sh}$. The large-scale fading coefficient  $\beta_{mk}$ from user $k$ and AP $m$ is given by
\vspace{-0.01cm}
			 {\small{   \begin{equation}\label{eq:pathloss}
			    \begin{aligned}
			    & \beta_{mk}= \left\{\begin{array}{ll}
			   & (-L-35\;\text{log}_{10}(r_{mk})  ) \times SF_{mk},\; \mbox{if  $r_{mk}>  d_1$,}  \\
			   & (-L-15\;\text{log}_{10}(d_1) -20\;\text{log}_{10}(r_{mk}))\times SF_{mk}\\
			   &  \qquad\qquad\qquad\qquad\qquad\qquad\mbox{if $d_0\leq r_{mk}\leq  d_1$, } \\
			   & ( -L-15\;\text{log}_{10}(d_1) -20\;\text{log}_{10}(d_0)) \times SF_{mk} \\
			  &  \qquad\qquad\qquad\qquad\qquad\qquad\qquad\mbox{if $ r_{mk}<  d_0$. } \\
			    \end{array}
			    \right.\\
			    \end{aligned}
			    \end{equation}}}where $SF_{mk}$ denotes the log-normal shadow fading between user $k$ and AP $m$ and  $	L$ is a constant depending on the carrier frequency,  the user and AP heights. All other simulation parameters are summarized in  Table \ref{parameters}. The performance of the proposed scheme is compared with the approach provided in \cite{CF_SC} but with uniform power allocation (we  refer to this  scheme as conventional CF).
			\begin{center}
				\footnotesize{
				\begin{tabular}{|c|l|c|l|}
					\hline
				Carrier frequency & $1.9 \;\text{GHz}$ & Bandwidth & $20 \; \text{MHz}$\\
					\hline
					Noise figure & $ 9 \; \text{dB} $ & AP antenna height & $15 \;\text{m}$\\
						\hline
						$\sigma_{sh}$& $8 \;\text{dB} $ & User antenna height & $1.7 \;\text{m}$\\
					\hline
						$T_c$& $   200   $ & $\rho_d, \rho_p$  & $200, 150 \;\text{mW} $\\
						\hline
				\end{tabular}
								\captionof{table}{Simulation parameters}\label{parameters} 
				}
			\end{center}

			\begin{figure}[!]
			\centering	
			\includegraphics[scale=.51]{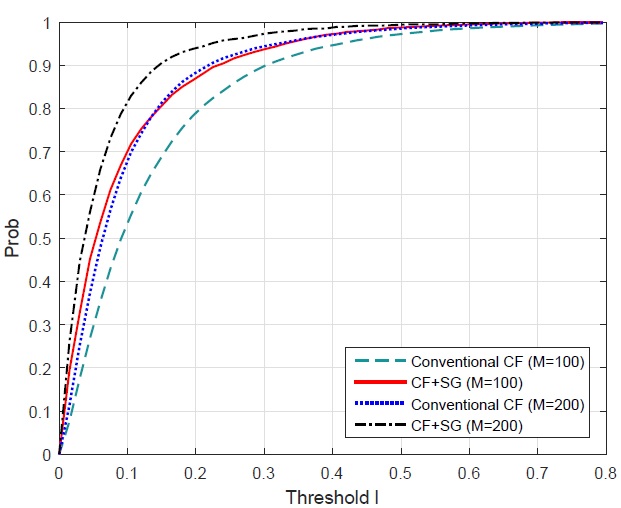}
			\captionsetup{font=footnotesize}
			\caption{Comparison of CDFs of normalized large-scale fading correlation for  $K=20,\alpha= 6, C=4$   \label{fig:CDF_coorelation} } 
		\end{figure}
		
			Figure \ref{fig:CDF_coorelation} shows the impact of the proposed spatial grouping (SG) on the normalized large-scale fading correlation for different values of  $M$.
		We can see that appropriate spatial user grouping  substantially reduces spatial correlation within each group which results in more favorable propagation. From Figure \ref{fig:CDF_coorelation}, we observe that  SG enables to achieve, for $M=100$, spatial correlation levels comparable to conventional CF with $M=200$. This means that SG can be a very practical alternative to  network densification.

\begin{figure}[!]
			\centering	
			\includegraphics[scale=.51]{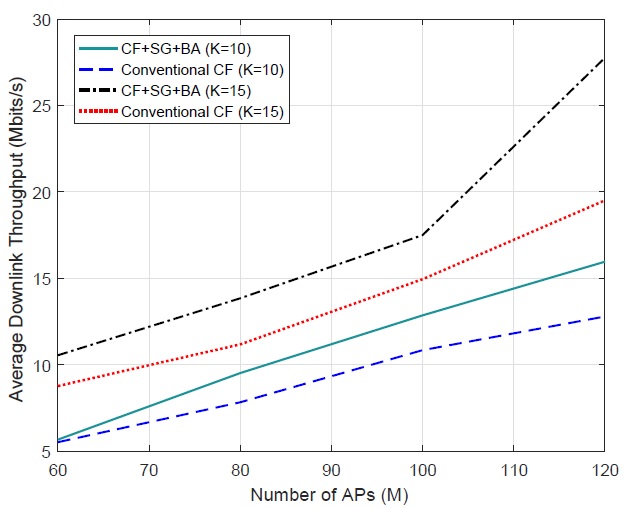}
			\captionsetup{font=footnotesize}
			\caption{Average Downlink Throughput versus the number of APs and different K values ($\tau=K$)   \label{fig:AVG_SRD} }
		\end{figure}

	Figure \ref{fig:AVG_SRD} exemplifies the average downlink net throughput versus the number of APs ($M$) for  $K=10$ and $15$,  respectively. 
	Figure \ref{fig:AVG_SRD} demonstrates that the performance of both approaches improves as M increases. This is  a direct consequence of  array gain increase. In addition,  Figure \ref{fig:AVG_SRD} also shows that the proposed SG with bandwidth allocation (BA) outperforms the conventional CF system. This is due to the fact that SG improves favorable propagation within each group. Indeed,   SG enables gains of  $18.5\%$  and $17.05\%$, for  $(M=100, K =10)$ and  $(M=100, K =15)$, respectively. Consequently, appropriate user grouping and selection can be a more practical and cost efficient alternative to  increasing the number of APs. Indeed,  for $K=15$, SG achieves approximately the same downlink throughput ( $14.956 \; Mbits/s$) at $M=85$ as the conventional CF system with $M=100$. 

			\section{Conclusion}
			
			In this work, we   investigated how favorable propagation can be  improved in  CF massive MIMO systems. We formulated an NP-hard spatial  user grouping problem based on  large-scale fading coefficients. We  designed an  SDR-based algorithm to find an efficient  sub-optimal solution. Orthogonal  frequency resources are also allocated to the spatial  groups. The simulation results showed that reducing  spatial  correlation between  active users through  efficient  spatial  grouping  considerably improves favorable  propagation, increases the  average throughout and enables to  achieve better performance with lower numbers of APs.


\begin{thebibliography}{50}
		\bibitem{CF_LSFD}	
		E. Nayebi, A. Ashikhmin, T. L. Marzetta, and H. Yang, \emph{Cell-Free
			Massive MIMO systems}, in 2015 49th Asilomar Conference on Signals,
		Systems and Computers, Nov 2015, pp. 695-699.
		\bibitem{CF_SC}
		H. Q. Ngo, A. Ashikhmin, H. Yang, E. G. Larsson, and T. L. Marzetta,
		\emph{Cell-Free Massive MIMO Versus Small Cells}, IEEE Transactions on
		Wireless Communications, vol. 16, no. 3, pp. 1834-1850, March 2017.
		\bibitem{Nayebi_TWC}
		E. Nayebi, A. Ashikhmin, T. L. Marzetta, H. Yang, and B. D. Rao, \emph{Precoding and Power Optimization in Cell-Free Massive MIMO Systems},
		IEEE Transactions on Wireless Communications, vol. 16, no. 7, pp. 4445-4459, July 2017.
		\bibitem{fp_massive}
		H. Q. Ngo, E. G. Larsson, and T. L. Marzetta, \emph{Aspects of favorable propagation
			in Massive MIMO}, in 2014 22nd European Signal Processing
		Conference (EUSIPCO), Sept 2014, pp. 76-80.
		\bibitem{DBLPzheng}
			Z. Chen and E. Bjornson, \emph{Channel hardening and favorable propagation in cell-free massive MIMO with stochastic geometry}, CoRR, 2017.
			\bibitem{Luo_SDR}
			Z. Q. Luo, W. K. Ma, A. M. C. So, Y. Ye, and S. Zhang, \emph{Semidefinite
				Relaxation of Quadratic Optimization Problems}, IEEE Signal Processing
			Magazine, vol. 27, no. 3, pp. 20-34, May 2010.
			\bibitem{EE_CF}
			H. Q. Ngo, L. N. Tran, T. Q. Duong, M. Matthaiou, and E. G. Larsson,
			\emph{On the Total Energy Efficiency of Cell-Free Massive MIMO}, IEEE
			Transactions on Green Communications and Networking, vol. PP, no. 99, 2017.
			\bibitem{prob_ineq}
			W. Hoeffding, \emph{Probability Inequalities for Sums of Bounded Random
				Variables}, Journal of the American Statistical Association, vol. 58, no.
			301, pp. 13-30, 1963.
			\bibitem{hard1}
			J. R. Correa and N. Megow, \emph{Clique partitioning with value-monotone
				submodular cost}, Discrete Optimization, vol. 15, pp. 26-36, 2015.
			\bibitem{Vandenberghe_1996}
			L. Vandenberghe and S. Boyd, \emph{Semidefinite programming}, SIAM Rev.,
			vol. 38, no. 1, pp. 49-95, March 1996.
			\bibitem{Stephen2004}
			S. Boyd and L. Vandenberghe, \emph{Convex Optimization}. Cambridge, U.K.:
			Cambridge Univ. Press, 2004.
			\bibitem{cvx}
			M. Grant and S. Boyd, \emph{CVX: Matlab software for disciplined convex
				programming, version 2.1}, http://cvxr.com/cvx, March 2014.	
		\end{thebibliography}
\end{document}